\documentstyle[twocolumn]{mn}

\input psfig

\def\beq{\begin{equation}}
\def\eeq{\end{equation}}
\def\bey{\begin{eqnarray}}
\def\eey{\end{eqnarray}}

\def\TJ{{T_{\rm J}}}
\def\TS{{T_{\rm S}}}
\def\TU{{T_{\rm U}}}
\def\TN{{T_{\rm N}}}

\def\and{{\rm and\ }}
\def\be{\begin{equation}}
\def\ee{\end{equation}}

\def\plottwob#1#2{\centering \leavevmode \epsfxsize=4cm \epsfbox{#1}
\hfil \epsfxsize=4cm \epsfbox{#2}}

\def\spose#1{\hbox to 0pt{#1\hss}}
\def\lta{\mathrel{\spose{\lower 3pt\hbox{$\sim$}}
    \raise 2.0pt\hbox{$<$}}}
\def\gta{\mathrel{\spose{\lower 3pt\hbox{$\sim$}}
    \raise 2.0pt\hbox{$>$}}}

\input epsf

\title[Simulations of Centaurs II: Individual Objects]
      {Simulations of the Population of Centaurs II: Individual Objects}

\author[J. Horner, N.W. Evans \& M.E. Bailey]
       {J. Horner$^{1,2}$, N.W. Evans$^{2,3}$ \& M.E. Bailey$^4$\\
       $^1$ Physikalisches Institut, Universit\"at Bern, Sidlerstrasse
       5, CH-3012 Bern, Switzerland \\
       $^3$ Theoretical Physics, Department of Physics, 1 Keble Rd,
       Oxford OX1 3NP\\
       $^3$ Institute of Astronomy, Madingley Rd, Cambridge, CB3
       0HA\\
       $^4$ Armagh Observatory, College Hill, Armagh, BT61 9DG,
       Northern Ireland.}
\date{}
\pagerange{\pageref{firstpage}--\pageref{lastpage}}
\pubyear{}

\begin{document}
\maketitle
\label{firstpage}

\begin{abstract}
Detailed orbit integrations of clones of five Centaurs -- namely, 1996
AR20, 2060 Chiron, 1995 SN55, 2000 FZ53 and 2002 FY36 -- for durations
of $\sim 3$ Myr are presented. One of our Centaur sample starts with
perihelion initially under the control of Jupiter (1996 AR20), two
start under the control of Saturn (Chiron and 1995 SN55) and one each
starts under the control of Uranus (2000 FZ53) and Neptune (2002 FY36)
respectively.  A variety of interesting pathways are illustrated with
detailed examples including: capture into the Jovian Trojans, repeated
bursts of short-period comet behaviour, capture into mean-motion
resonances with the giant planets and into Kozai resonances, as well
as traversals of the entire Solar system.  For each of the Centaurs,
we provide statistics on the numbers (i) ejected, (ii) showing
short-period comet behaviour and (iii) becoming Earth and Mars
crossing. For example, Chiron has over $60 \%$ of its clones becoming
short-period objects, whilst 1995 SN55 has over $35 \%$. Clones of
these two Centaurs typically make numerous close approaches to
Jupiter.  At the other extreme, 2000 FZ53 has $\sim 2 \%$ of its
clones becoming short-period objects.  In our simulations, typically
$20 \%$ of the clones which become short-period comets
subsequently evolve into Earth-crossers.
\end{abstract}

\begin{keywords}
minor planets, asteroids -- planets and satellites: general --
celestial mechanics, stellar dynamics -- Kuiper belt
\end{keywords}

\section{Introduction}

The Centaurs are a transition population of minor bodies between the
trans-Neptunian objects and the Jupiter-family comets (see, for
example, Horner et al. 2003 and the references therein). Centaurs
typically cross the orbits of one or more of the giant planets and
have relatively short dynamical lifetimes ($\sim$$10^6$\,yr). Their
properties are exemplified by the first known Centaur, Chiron, which
was found in 1977 on Palomar plates (Kowal et al. 1979). Chiron is a
large minor body with perihelion close to or within the orbit of
Saturn and aphelion close to the orbit of Uranus.  The Centaurs have
so far largely eluded the attention of numerical integrators. The only
ones that have hitherto been the subject of detailed dynamical
investigations are Chiron itself (Hahn \& Bailey 1990, Nakamura \&
Yoshikawa 1993) and Pholus (Asher \& Steel 1993). Dones et al. (1996)
also looked briefly at four Centaurs, including Chiron and Nessus.
All these investigations were for durations of less than 1 Myr and
involved modest numbers of clones.

Horner et al. (2004, hereafter Paper I) integrated the orbits of
23\,328 clones of 32 selected Centaurs and used the dataset to
evaluate statistical properties of the Centaurs in a model Solar
system containing the Sun and the four giant planets. Hence, these
longer numerical integrations with large numbers of clones provide
better statistics and highlight some unusual past histories and future
fates for Centaurs. In this companion paper, the behaviour of clones
of five of these Centaurs -- namely, 1996 AR20, Chiron, 1995 SN55,
2000 FZ53 and 2002 FY36 -- are studied in more detail. The objects are
chosen to span a wide range of properties. 1996 AR20 has the shortest
half-life in our sample, while 2000 FZ53 has the longest
half-life. 1995 SN55 is the Centaur with the brightest absolute
magnitude (hence potentially the largest Centaur known), while Chiron
is the only one confirmed to display cometary out-gassing.

Horner et al. (2003) introduced a new classification system for
cometary-like bodies according to the planets under whose control the
perihelion and aphelion lie.  For example, we classify Chiron as an SU
object, by which we mean that the position of its perihelion lies
within Saturn's zone of control, and that the position of its aphelion
lies within Uranus' zone of control. It is apparent that perturbations
at perihelion, by Saturn, will act primarily to move the position of
the aphelion, and vica-versa. In other words, the motion near Saturn
determines whether or not the body gets to Uranus, or is captured to a
more tightly bound orbit, or expelled.  Conversely, perturbations by
Uranus, near aphelion, largely determine the future perihelion
distance.  So, in a wider sense, Saturn also `controls' the aphelion
(and Uranus the perihelion), as it determines its numerical value.
However, in this paper, whenever we talk of a planet controlling a
minor body at perihelion (or aphelion), we mean that the motion at
perihelion (or aphelion) lies in the zone of control of that planet.

For our selected 5 Centaurs, there is one object with perihelion under
the control of Jupiter (1996 AR20), two under the control of Saturn
(Chiron and 1995 SN55), and one each under the control of Uranus (2000
FZ53) and Neptune (2002 FY36). Clones of the objects were created by
incrementally increasing (and decreasing) the semi-major axis $a$ of
the object by 0.005 au, the eccentricity $e$ by 0.005, and the
inclination $i$ by $0.01^\circ$. Nine values were used for each of
these elements, with the central (fifth) value of the nine having the
original orbital elements for the Centaur, as taken from {\it The
Minor Planet Center}. The other orbital elements aside from $a$, $e$
and $i$ are unchanged (see Paper I for more details). This procedure
yielded 729 clones of each Centaur, all of which were numerically
integrated for up to 3 Myr. In this paper, we restrict ourselves to
just 2 particularly interesting clones for each Centaur.

Although all 5 of our selected Centaurs have reasonably reliable
ephemerides, only Chiron has been the subject of sustained interest
from observers.  For Chiron, there are long-term photometric studies
of the behaviour of the object (Dufford et al., 2002), detailed
analyses of its reflectance spectrum (Foster et al. 1999), as well as
the use of archival pre-discovery images of the object (Bus et al.,
2001). There are little observational data on the remaining four
objects.

The detailed studies of individual clones of these objects are
important to illustrate some of the dynamical pathways in the Solar
system.  Objects in very stable r\'egimes in the Solar system (such as
some resonances) are long-lived and could be potential targets for new
surveys. A good example is the possible long-lived belt of objects
between Uranus and Neptune claimed by Holman (1997). Objects in
unstable r\'egimes must evolve, and correlations between observables
and orbital properties are then expected.  For example, bluer colours
might indicate a younger, fresher surface and so be indicative of
recent cometary activity. So, a Centaur with blue colours (such as
Chiron) could be a candidate for a passage through a cometary phase in
the recent past. Individual examples allow us to match an orbital
history to such a presumed pathway.

The paper is organized according to object, with 1996 AR20 studied in
\S 2, Chiron in \S 3, 1995 SN55 in \S 4, 2000 FZ53 in \S 5 and 2000
FY36 in \S 6.

\begin{figure} 
\plottwob{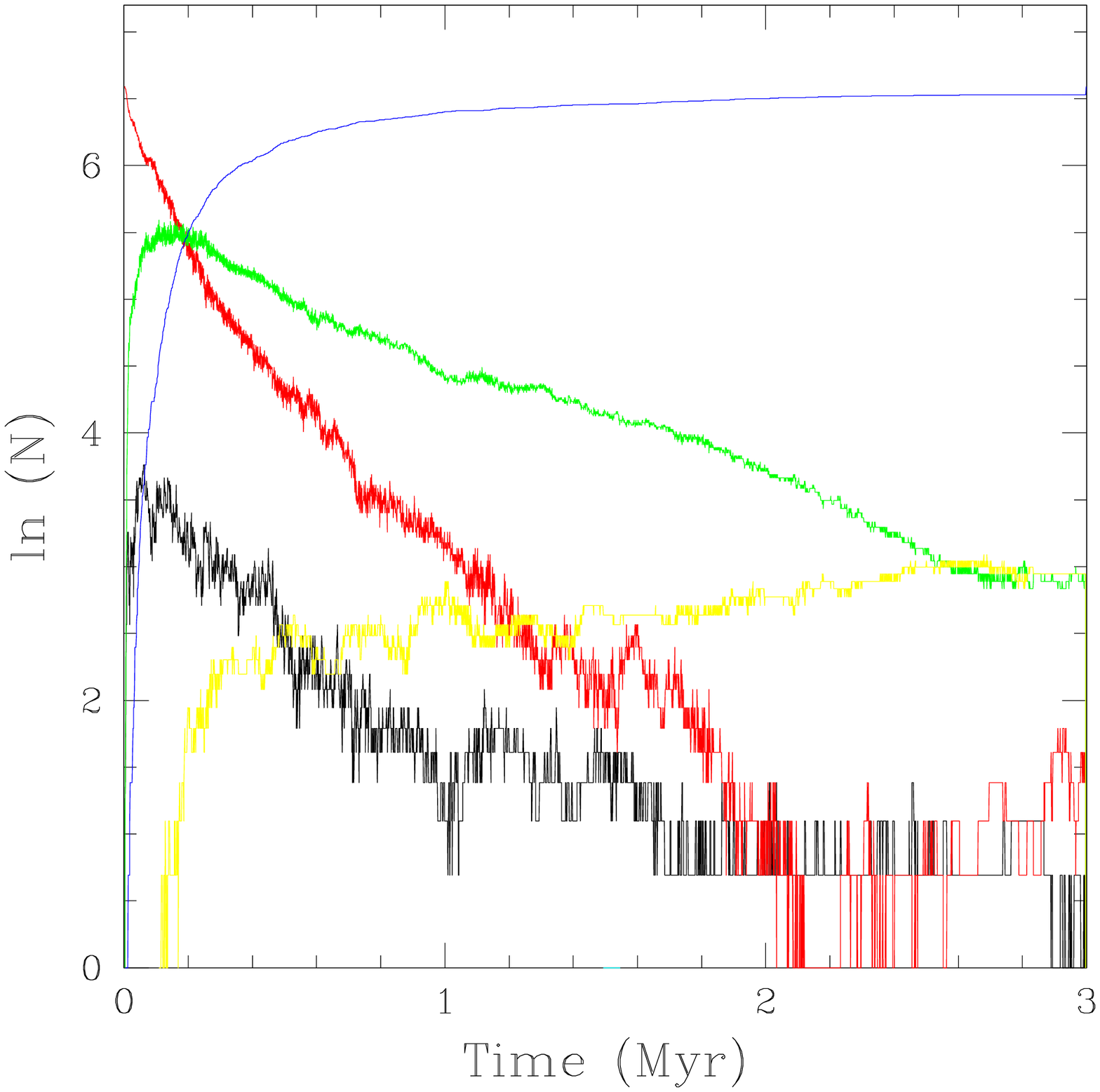}{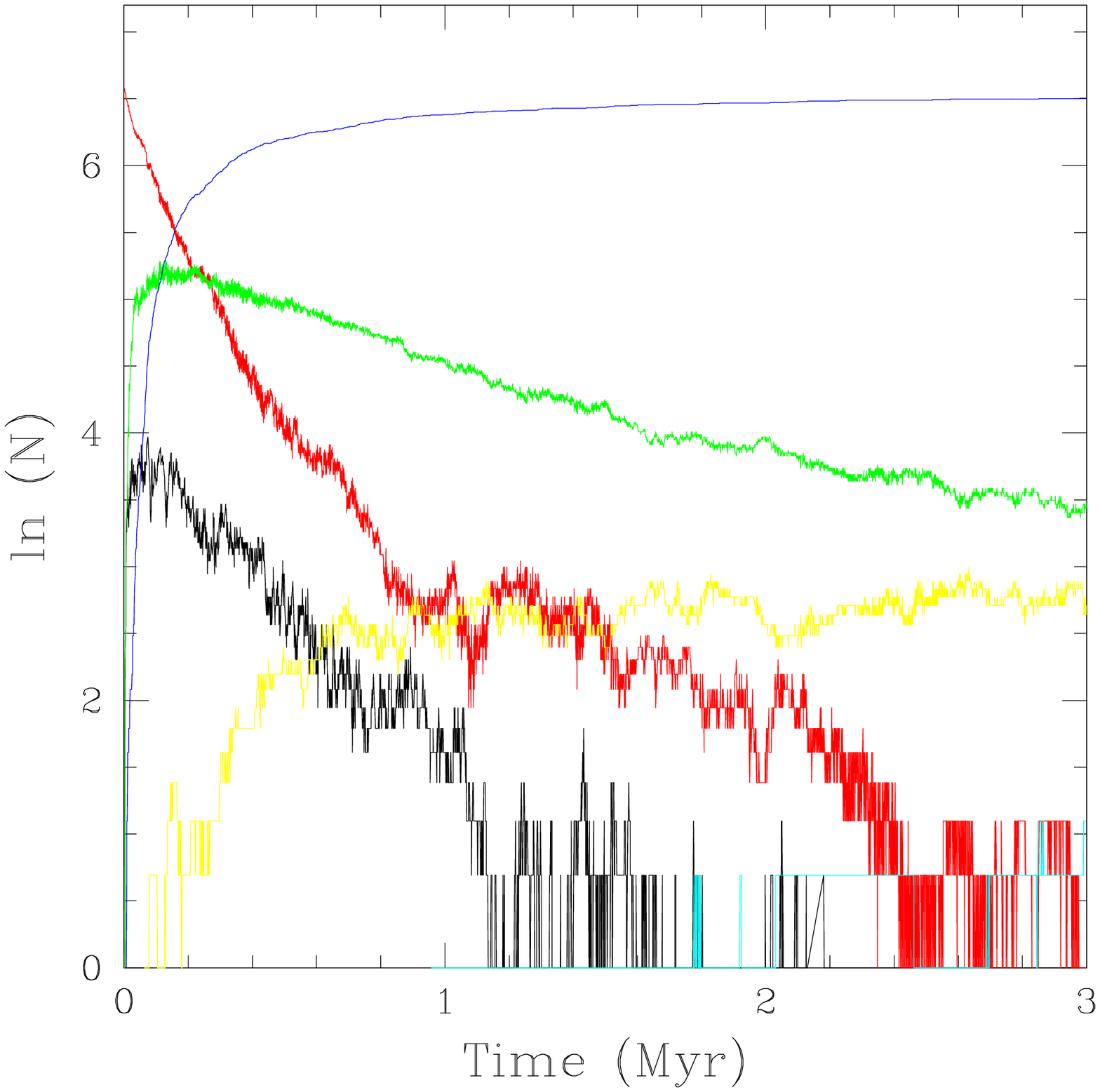} 
\caption[] {The evolution of the population of clones of 1996 AR20
subdivided according to the planet controlling the perihelion (objects
controlled by Jupiter, Saturn, Uranus and Neptune are red, green,
yellow and cyan respectively). Also shown are the evolution of the
number of short-period comets (black), trans-Neptunian objects and
ejected objects (blue). The left panel shows the results from the
forward integration, the right the backward integration.  [This colour
convention is employed in all following plots of this nature.]}
\label{fig:AR20classtime}
\end{figure} 
\begin{figure}
\epsfysize=10cm \centerline{\epsfbox{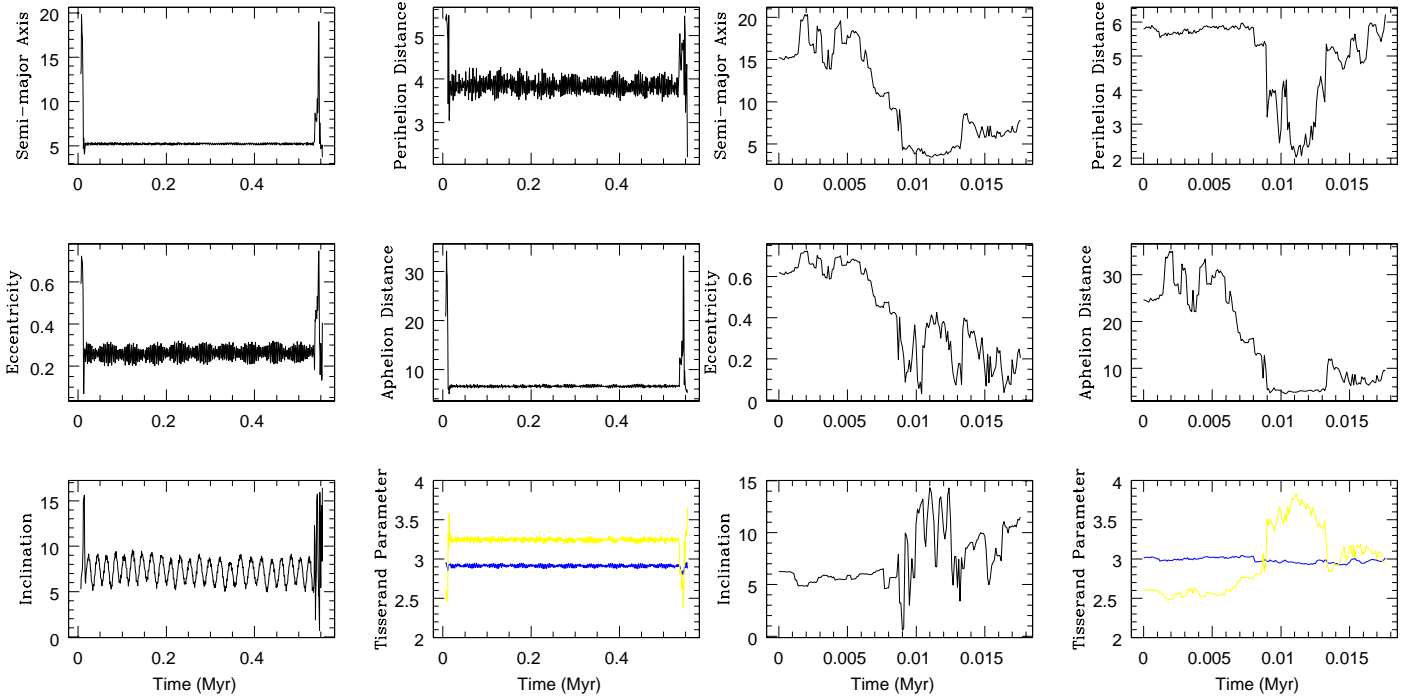}}
\caption[] {The evolution of the 12th clone of 1996 AR20 in the
forward direction. Sub-panels show the evolution of semimajor axis,
perihelion and aphelion distance (all in au), inclination (in degrees)
and eccentricity. In the plot of Tisserand parameter, the value of
$\TJ$ is plotted in blue and $\TS$ in yellow. This convention is
followed in all similar plots. Note that the clone is rapidly trapped
into a 1:1 mean motion resonance with Jupiter until ejection after
$\sim 0.5$ Myr.}
\label{fig:AR20a}
\end{figure}
\begin{figure}
\epsfysize=10cm \centerline{\epsfbox{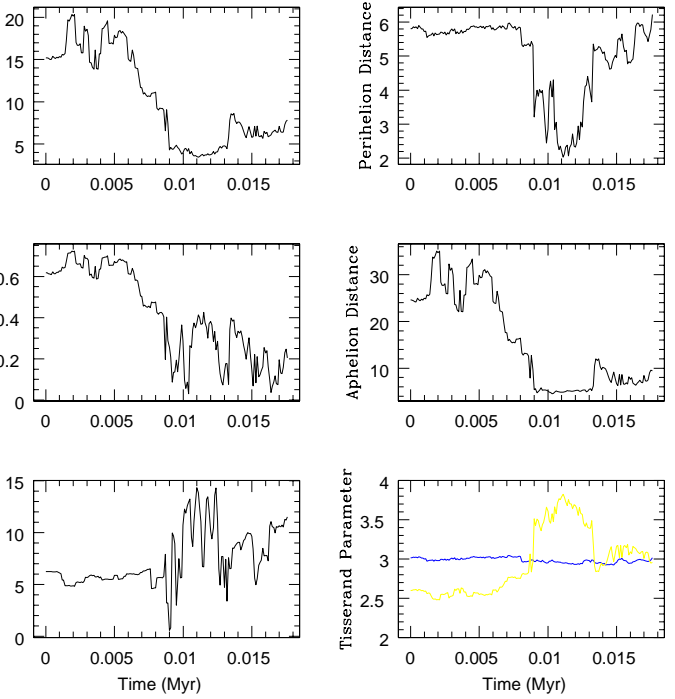}}
\caption[] {The evolution of the 66th clone of 1996 AR20 in the
forward direction. The clone hits the surface of Saturn after 18
kyr.}
\label{fig:AR20b}
\end{figure}

\section{Evolution of a JN object: 1996 AR20}

1996 AR20 is a JN object with its perihelion under the control of
Jupiter and its aphelion under the control of Neptune.  Among the
Centaurs, 1996 AR20 has the shortest known half-lives, namely 540 kyr
in the forward and 594 kyr in the backward direction. Its orbit is
interesting as its initial position lies close to two prominent
mean-motion resonances. The initial value of semi-major axis in the
integrations was 15.2 au, which is within 0.02 au of the 1:5
mean-motion resonance with Jupiter and within 0.06 au of the 1:2
mean-motion resonance with Saturn.  In addition, 1996 AR20 has an
eccentricity of 0.627 so that it can approach all the major outer
planets close enough to be perturbed. These factors all contribute
towards making 1996 AR20 one of the least stable Centaurs. Of the 729
clones, 62 become Earth-crossers, 154 become Mars-crossers and 340
become short-period comets in the forward integration. These numbers
are all slightly larger in the backward integration, namely 89, 194,
and 406 respectively.

Figure \ref{fig:AR20classtime} shows how the population of clones of
1996 AR20 changes over time. Initially, all 729 clones have perihelion
under the control by Jupiter, but by the end of the simulation, in
both the forward and backward directions, over 650 of the clones have
been ejected. The number of objects under the control of Jupiter
rapidly decays, with most either being ejected, or moving to the
control of Saturn, or transferred to cometary orbits. The numbers in
each of these classes peaks early within the simulation and then
decays as more and more objects are ejected. Only a small number of
clones of 1996 AR20 evolve so that the perihelion is under the control
of Uranus and Neptune. The great majority of objects are ejected by
either Jupiter or Saturn, giving very few the opportunity to evolve
all the way out to Neptune.

\subsection{A Source for Jovian Trojan Asteroids}

Figure \ref{fig:AR20a} shows the evolution of the 12th
clone\footnote{The clone label is useful for our internal data
management but carries no other physical meaning.} of 1996 AR20,
integrated in the forward direction. The initial semimajor axis,
eccentricity and inclination of this clone are $a = 15.177$ au, $e =
0.617$ and $i = 6.17^\circ$. The clone is rapidly captured into a 1:1
mean-motion resonance with Jupiter, which it then occupies for over
0.5 Myr before ejection from the Solar system. The clone displays
quite large variations in $a$, $e$ and $i$ whilst in the resonance. By
plotting the positions over time, it is clear that the clone follows a
tadpole orbit librating about the Lagrange point. This is significant
as it shows that Centaurs can be captured into the 1:1 resonance with
Jupiter. Hence, there may well be Jovian Trojans that were originally
Centaurs and vice versa.  It would be interesting to see whether any
Jovian Trojans display cometary out-gassing, since recently captured
Centaurs may still contain volatiles, whilst any Trojans captured from
an original Main Belt asteroidal orbit are unlikely to display such
activity.

In our Centaur orbital integrations, we find that clones are quite
frequently trapped into 1:1 mean-motion resonances with all the giant
planets.

\subsection{A Collision with Saturn}

Figure \ref{fig:AR20b} shows the behaviour of the 66th clone of 1996
AR20, whose initial orbital elements were $a = 15.177$ au, $e =
0.617$, $i = 6.23^\circ$ (almost the same as the 12th clone!). The
66th clone impacts upon Saturn at the end of its lifetime, 18 kyr
after the start of the integration.  In Paper I, we calculated that
Centaurs impact onto the surface of Saturn at a rate of 1 every 28
kyr.  The perihelion of the clone starts the simulation under the
control of Jupiter, and perturbations by this planet cause a number of
changes in the semi-major axis of the clone. Finally, a series of
close encounters reduce the perihelion and aphelion distances for the
object until it twice becomes a cometary body (at around 12 kyr, very
briefly, and then for a more prolonged period from 13 kyr to 15
kyr). After this, the clone's perihelion and aphelion distances
increase until the perihelion lies just beyond the orbit of Jupiter
and the aphelion lies under Saturn's control. The object finally
collides with Saturn at aphelion, roughly 18 kyr from the present.

\begin{figure} 
\plottwob{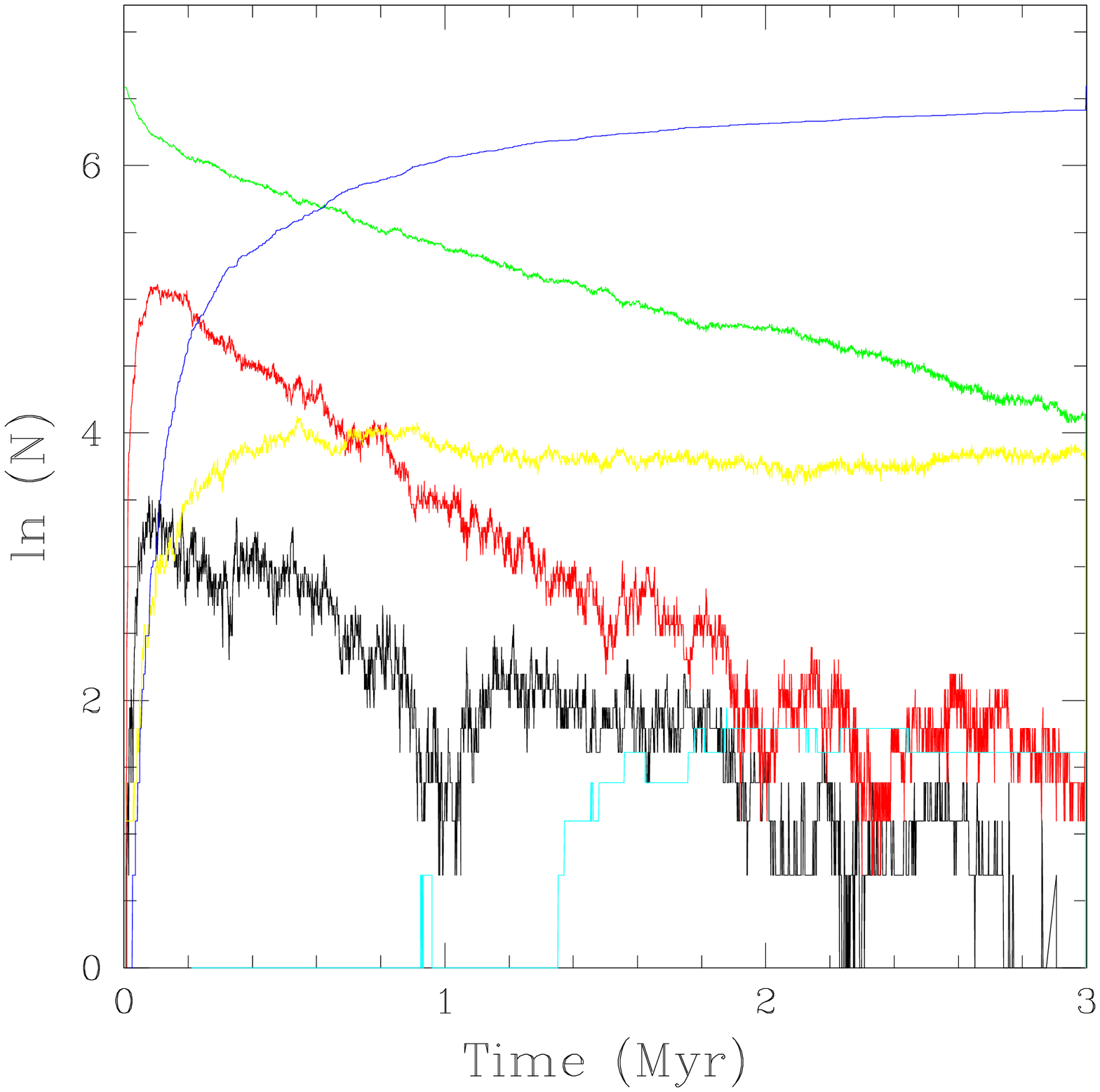}{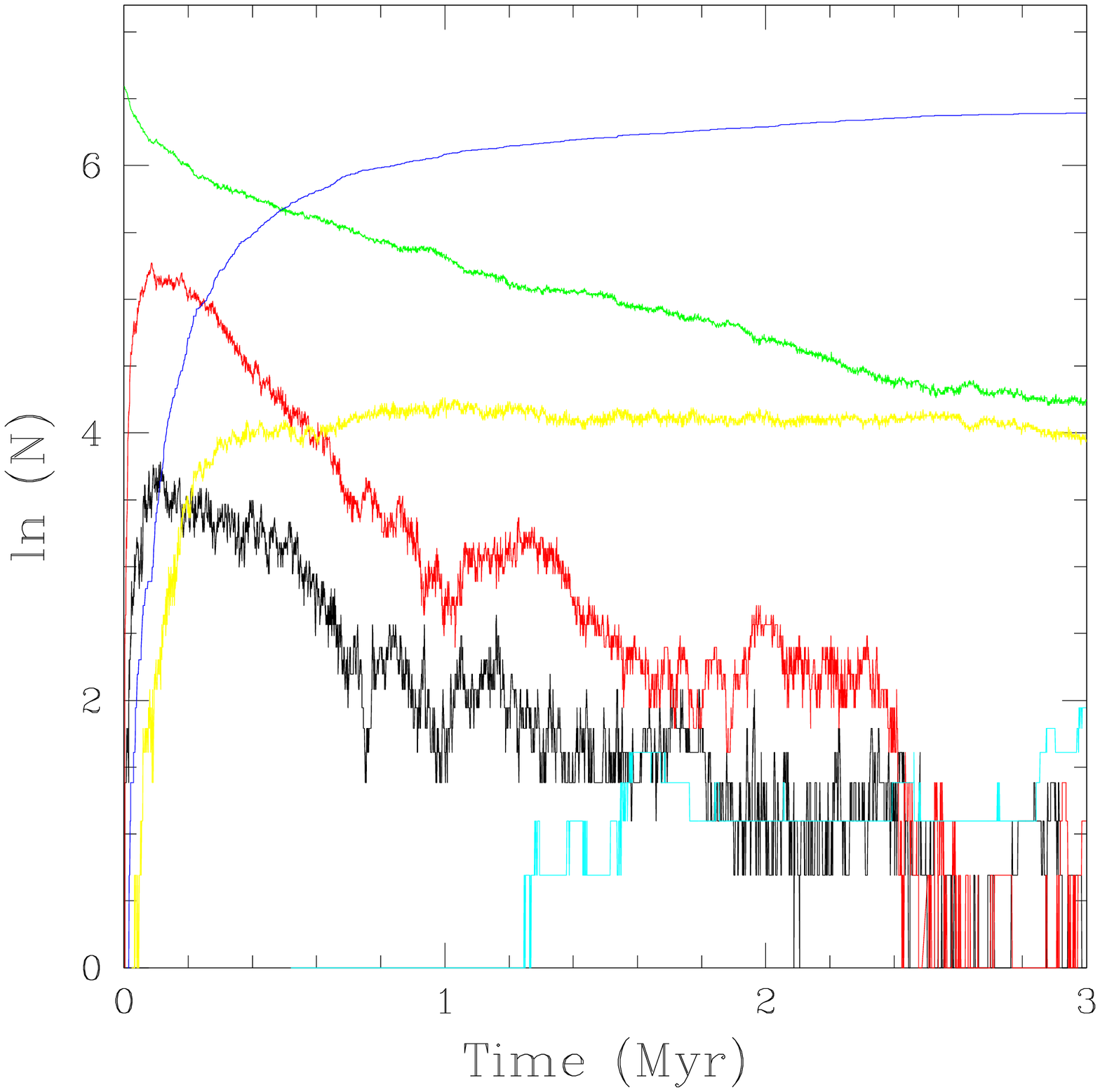} 
\caption[] {The numbers of clones of Chiron controlled by Jupiter
(red), Saturn (green), Uranus (yellow) and Neptune (cyan), together
with the numbers of short-period comets (black), trans-Neptunian and
ejected objects (blue), plotted against time. The left (right) panel
shows the results from the forward (backward) integration.}
\label{fig:Chironclasstime}
\end{figure} 
\begin{figure}
\epsfysize=10cm \centerline{\epsfbox{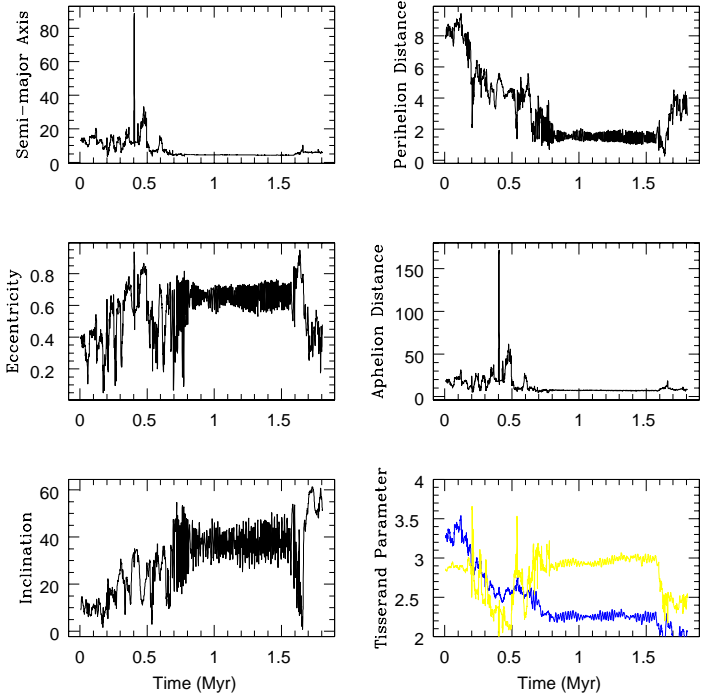}}
\caption{The evolution of the 206th clone of Chiron in the forward
direction.  and eccentricity. In the plot of Tisserand parameter, the
value of $\TJ$ is plotted in blue and $\TS$ in yellow.  Note the
prolonged spell ($\sim 1$ Myr) as a short-period comet.}
\label{fig:Chironb}
\end{figure}
\begin{figure}
\epsfysize=10cm \centerline{\epsfbox{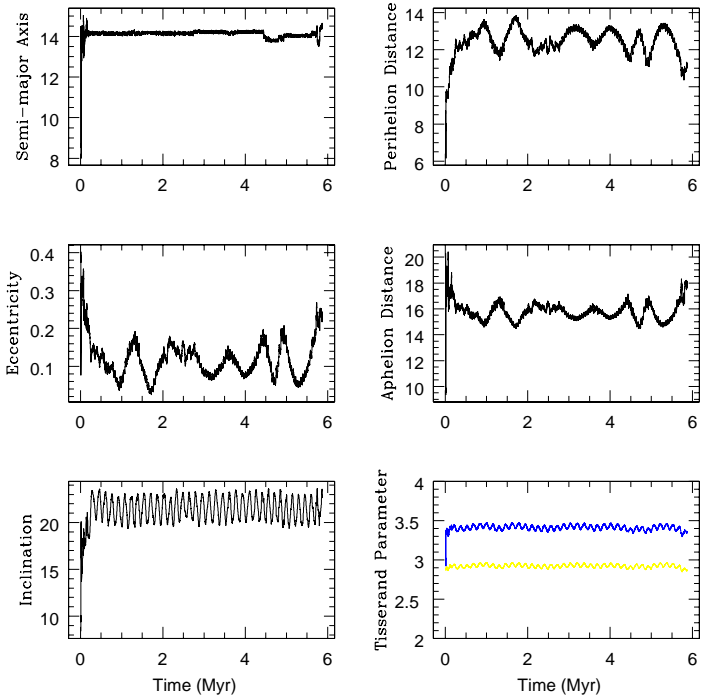}}
\caption{The evolution of the 78th clone of Chiron in the forward
direction. Note the stable, nearly constant behaviour of the orbital
elements as the clone is transferred to a long-lived orbit.}
\label{fig:Chirona}
\end{figure}

\section{Evolution of a SU object: Chiron}

Chiron was the first Centaur to be discovered in 1977. Pre-discovery
images allow the orbit to be traced all the way back to the perihelion
passage of 1895 (see Kowal et al. 1979). Chiron has a coma which
undergoes variations in brightness (Meech \& Belton 1989, Luu \&
Jewitt 1990). Chiron's photometric activity is sporadic and apparently
unrelated to heliocentric distance (Duffard et al. 2002). For example,
the increase in brightness during 1988-1991 (e.g., Tholen et al. 1988)
was followed by a period of minimal activity as the object passed
through perihelion in 1996. Also unusual is the size of Chiron -- with
an absolute magnitude $H$ of 6.5, it is one of the largest Centaurs
(only Chariklo and 1995 SN55 are brighter). The object has a half-life
of 1.03 Myr (forwards) and 1.07 Myr (backwards).  In the forward
simulation, 415 objects become short-period objects, of which 84
become Earth crossing and 180 become Mars crossers. In the backward
simulation, these numbers are slightly larger at 445, 110 and 208
respectively. In other words, significantly more than half of the
clones become short-period comets at some point within their
evolution, suggesting that it is quite likely that Chiron has been a
cometary object at some point in the past and may well become one
again in the future. This ties in well with the work of Hahn \& Bailey
(1990), although they found a much greater discrepancy between the
likelihoods of the object being a short-period comet in the future and
in the past.  Figure \ref{fig:Chironclasstime} shows how the overall
population of clones of Chiron changes over time. Over the period of
the integration, around 600 clones are ejected in both the forward and
backward integrations.

\subsection{A Long-Lived Short-Period Comet}

Figure \ref{fig:Chironb} shows the evolution of the 206th clone of
Chiron, which started the integrations with $a = 13.591$ au, $e =
0.394$ and $i = 6.90^\circ$. This clone displays short-period cometary
behaviour for almost 1 Myr. During the early part of the evolution,
encounters act to reduce its perihelion distance so that it comes
under Jupiter's control. Once this happens, the behaviour of the
object becomes more chaotic, leading to a near-ejection at 400 kyr,
together with a number of short spells as a short-period comet
(e.g. at 200 kyr). Finally, at around 700 kyr, the object is
transferred into a cometary orbit of short-period. At 800 kyr, the
object is captured into an orbit close to the 6:5 mean-motion
resonance with Jupiter and the 3:1 mean-motion resonance with
Saturn. After around 50 kyr in this resonance, the semi-major axis of
the clone is reduced to slightly less than 4.5 au, and the object
enters the 5:4 mean-motion resonance with Jupiter, which it occupies
for approximately 350 kyr. After this time, the semi-major axis
gradually decreases to smaller and smaller values, until at around 1.3
Myr the object enters the 4:3 resonance with Jupiter. It leaves this
resonance briefly at the 1.4 Myr mark, but then re-enters it at around
1.45 Myr. Over all this time, the eccentricity and inclination of the
clone experience rapid oscillations, with the inclination at times
reaching over $50^\circ$. The perihelion and aphelion values also
oscillate wildly, although the object only becomes Earth-crossing at
the end of its time as a short-period comet. Shortly after this,
encounters with Jupiter act to raise the perihelion distance slightly
and eject it from the Solar system. The amount of time spent as a
cometary object for this clone, at $\sim 1$ Myr, is fairly
exceptional. However, it does show that there is the scope for
Centaurs to be captured into short-period cometary orbits for very
long periods. As Hahn \& Bailey (1990) first emphasised, this is
interesting and worrisome -- an object the size of Chiron occupying a
short-period cometary orbit for this period of time would pollute the
inner Solar system with huge amounts of dust and debris. Though such
long-term captures are uncommon, they are not by any means unusual
within our sample of clones.

\subsection{A Very Stable Resonant Orbit}

Figure \ref{fig:Chirona} shows the orbital evolution of the 78th clone
of Chiron, which had initial orbital elements of $a = 13.581$ au, $e =
0.384$ and $i = 6.94^\circ$. This clone is almost immediately captured
into a very stable orbit, at around a semi-major axis of 14.15 au, in
which it remains for the duration of the 3 Myr integration. The 5:9
resonance with Saturn lies at $\sim 14.2$ au. Of course, the 2:9
resonance with Jupiter also lies at roughly the same location, but its
effect is likely to be weaker on account of the high order of the
resonance and the large perihelion distance of the clone.  It is
interesting that, during the period of stable behaviour, the
inclination displays very smooth cyclical variations, between around
$20^\circ$ and $24^\circ$, whilst the eccentricity (and hence the
perihelion $q$ and aphelion $Q$ distances) display variations which
are much less regular. Throughout the stable period, the clone has a
low eccentricity and hence orbits entirely between Saturn and Uranus.
This is another illustration of the point made by Holman (1997) and
Evans \& Tabachnik (1999), namely that low eccentricity orbits between
the planets can be very stable.

\begin{figure} 
\plottwob{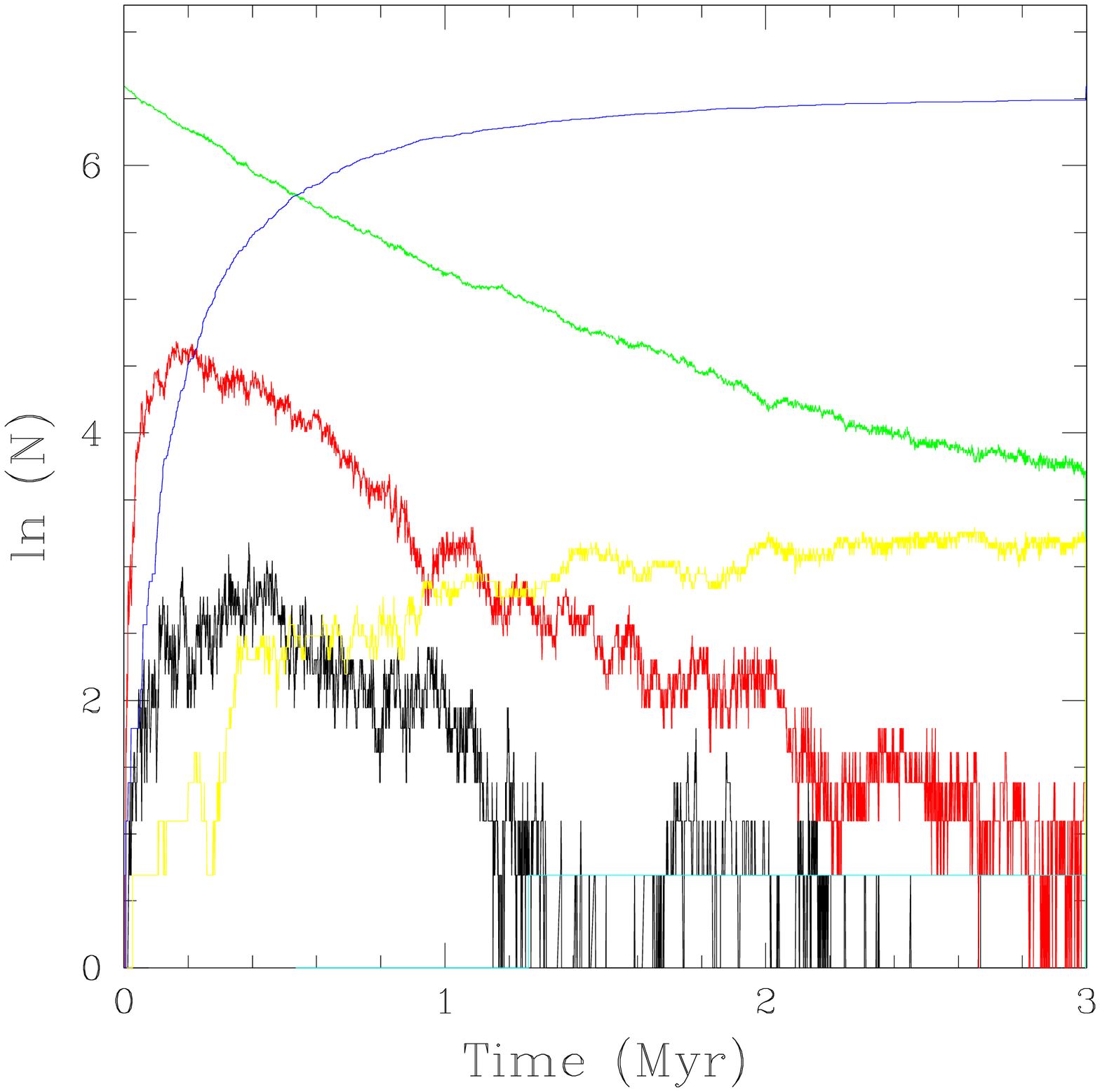}{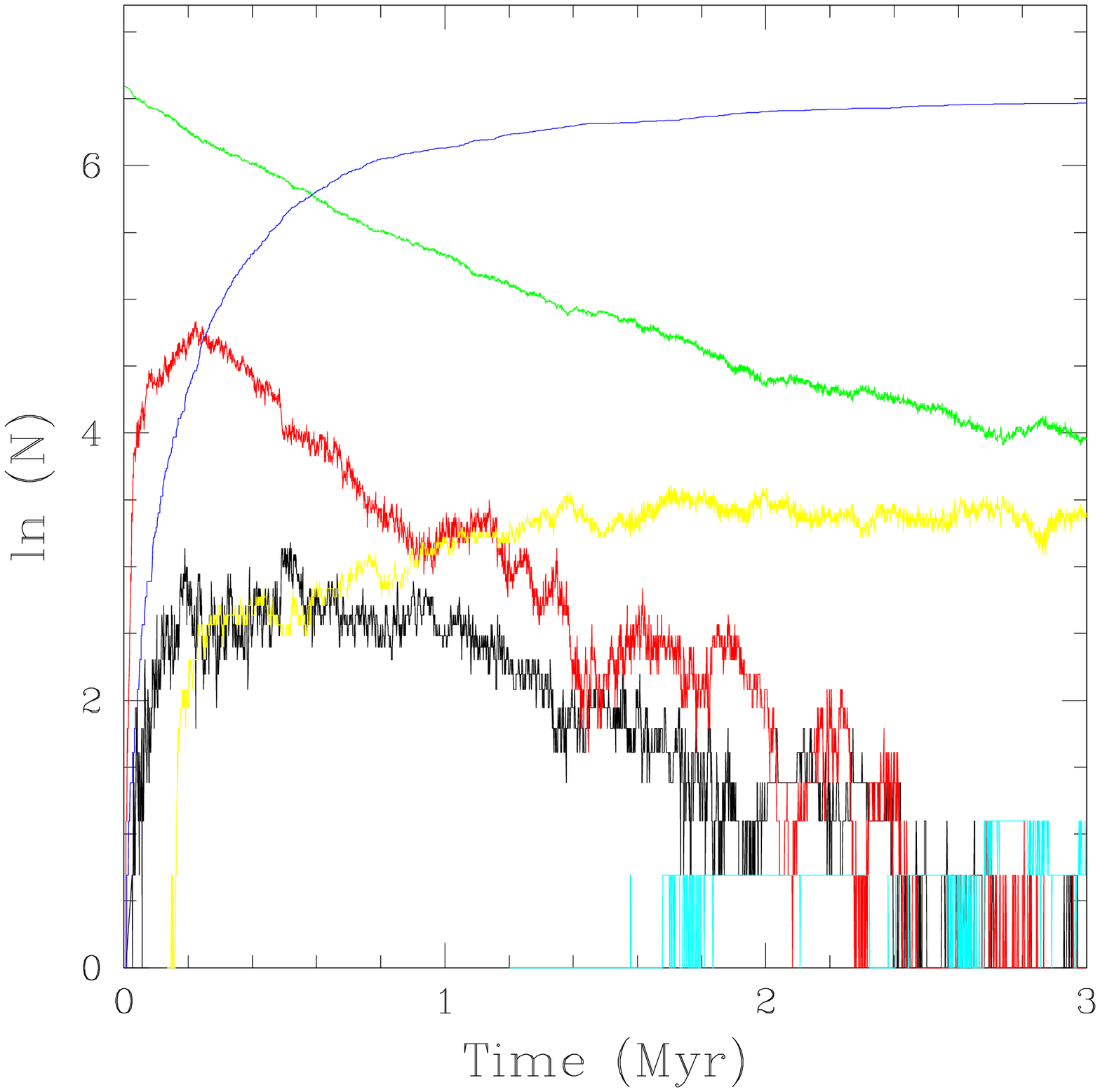} 
\caption[] {The numbers of clones of 1995 SN55 controlled by Jupiter
(red), Saturn (green), Uranus (yellow) and Neptune (cyan), together
with the numbers of short-period comets (black), trans-Neptunian and
ejected objects (blue), plotted against time. The left (right) panel
shows the results from the forward (backward) integration.}
\label{fig:SN55classtime}
\end{figure} 
\begin{figure}
\epsfysize=10cm \centerline{\epsfbox{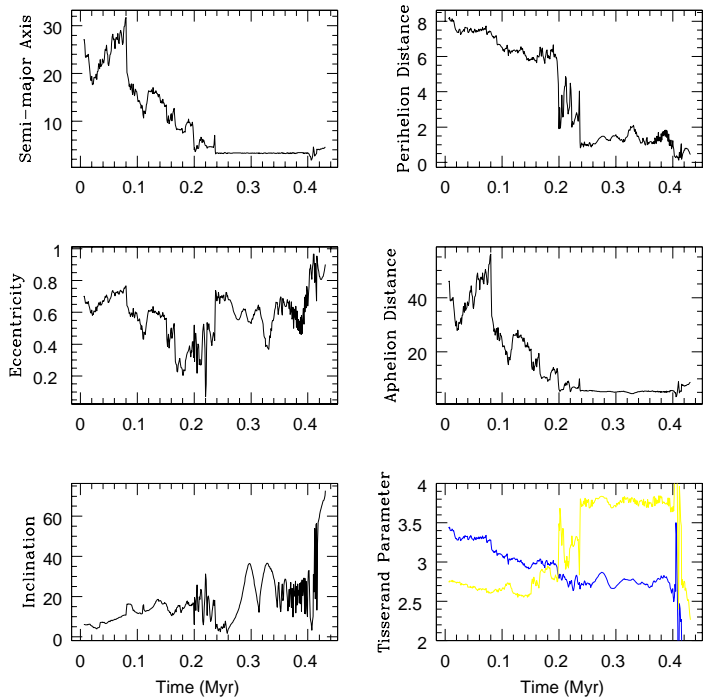}}
\caption{The behaviour of the 103rd clone of 1995 SN55 in the forward
direction. In the plot of Tisserand parameter, the value of $\TJ$ is
plotted in blue and $\TS$ in yellow.  Note the spell as a short-period
comet, Earth-crosser and finally Sun-grazer.}
\label{fig:SN55a}
\end{figure}
\begin{figure}
\epsfysize=10cm \centerline{\epsfbox{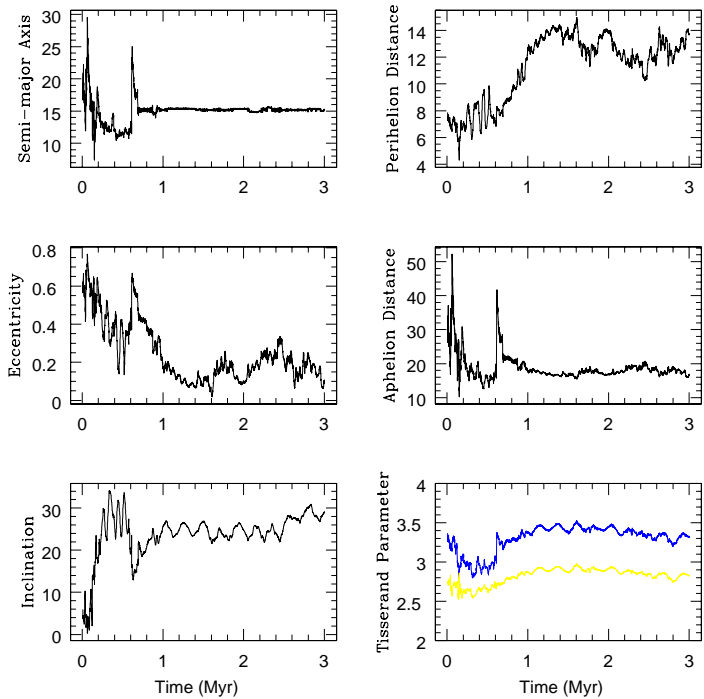}}
\caption{The orbital evolution of the 160th clone of 1995 SN55 in the
forward direction. Note that the endpoint of the clone's evolution is
an orbit that lies almost entirely between Saturn and Uranus and is
quite stable. }
\label{fig:SN55b}
\end{figure}

\section{Evolution of a SE object: 1995 SN55}

1995 SN55 is an intriguing object that is surely worthy of more study
from observers -- if only in the first instance to recover it! It is
only known from observations covering an arc of 36 days.  According to
its absolute magnitude ($H=6.0$), 1995 SN55 is the largest of the
Centaurs, with a diameter somewhere between 170 and 380 km (see e.g.,
Table 2 of Paper I). In addition, it has a high eccentricity ($e =
0.663$), which means that at perihelion the object lies 7.9 au from
the Sun, whilst at aphelion, it reaches out to 39.2 au. The half-life
is 0.701 Myr in the forward and 0.799 Myr in the backward direction.
In the forward integration, 250 of the initial 729 objects became
short-period comets at some point, with 50 becoming Earth-crossing and
112 becoming Mars-crossing. In the backward integration, these numbers
are slightly larger at 278, 55 and 118 respectively.  Figure
\ref{fig:SN55classtime} shows how the overall population of clones of
1995 SN55 changes over time. The unstable nature of this object is
shown clearly in the rapid rate at which clones are ejected. In both
forward and backward integrations around 650 of the clones are removed
from the simulations by their end at 3 Myrs.

\subsection{A Source for Earth-Crossers}

Figure \ref{fig:SN55a} shows the evolution of the 103rd clone of 1995
SN55. This clone has initial orbital parameters of $a = 23.549$ au, $e
= 0.658$ and $i = 4.98^\circ$. It spends a prolonged spell of time as
a short-period object, during which it approaches the Earth's orbit
closely and actually becomes Earth-crossing near the end of its
short-period lifetime.  In its early evolution, the clone experiences
a number of changes in semi-major axis, due mainly to encounters with
Saturn around perihelion. A particularly close encounter with Saturn
at around 90 kyr reduces the aphelion distance $Q$ from $\sim 60$ au
to below 30 au. This encounter is visible as a clear discontinuity in
the plots for $a$, $e$ and $Q$. After this, perturbations act
systematically to reduce the perihelion distance, until the object
enters Jupiter's sphere of control. Then, a deep encounter at Jupiter
reduces the perihelion distance still further to $\sim 2$ au. For a
further 40 kyr, the object moves on a chaotic orbit controlled by
Jupiter, until at around 240 kyr, it is captured into a 2:1
mean-motion resonance with Jupiter in which it resides for $\sim 150$
kyr. This is almost identical to the 5:1 mean-motion resonance with
Saturn. The rough 5:2 commensurability of Jupiter and Saturn clearly
plays an important role in providing a pattern of stable niches in
which objects can survive for long periods of time.

At the beginning of the stay in the resonance, the eccentricity and
inclination of the clone (and hence the perihelion $q$ and aphelion
$Q$ distances) vary erratically. After $\sim 20$ kyr, they become more
stable, and start to display gradual, long term variations. This is
most obvious in the inclination of the clone, which is gradually
pumped from a few degrees to two peaks of around $36^\circ$.  After
some 340 kyr, the behaviour of $e$ and $i$ begins to cycle far more
rapidly, leading to equally rapid fluctuations in the behaviour of $q$
and $Q$. Finally, at around 400 kyr, the clone leaves the mean-motion
resonance, and moves inwards to become Earth-crossing. Towards the end
of its life, the clone becomes Sun-grazing. However, the simulation is
not trustworthy at very small $q$, owing to the fixed time step of 120
days (see Paper I). Nonetheless, it would be interesting to understand
the effects of the impact of $\sim 100$ km sized Centaurs (like 1995
SN55) on the Sun itself, for example, in terms of enhanced metallicity
and increased reconnection effects.

\subsection{A Stable End-Point in the Outer Solar System}

Figure \ref{fig:SN55b} shows the behaviour of the 160th clone of 1995
SN55, which has initial orbital elements of $a = 23.549$ au, $e =
0.673$ and $i = 5.04^\circ$. During the first 400 kyr of the evolution
of this clone, it undergoes significant changes in its orbit, due
mainly to the effects of Jupiter and Saturn. At one point (just before
the 200 kyr mark), the clone's perihelion dips very briefly to a mere
4 au, before rising again. Whilst the object is being perturbed in
this way, it experiences a fairly rapid rise in inclination, from the
initial value of around $4^\circ$ to a peak over $30^\circ$.  After
400 kyr, the changes in the orbital elements of the clone become less
severe, with the exception of one large `kick' given at perihelion by
Saturn, just after the 600 kyr mark. Shortly after this, the object
falls into a stable orbit with $a \approx 15$ au. It then spends the
remainder of the integration in this region of the Solar system. For
the bulk of its stay, the clone has a semi-major axis of between 15
and 15.5 au. In this region, there are a number of mean-motion
resonances which may be important in the behaviour of this
clone. First, the 1:5 resonance with Jupiter lies at about 15.22 au,
and the 1:2 resonance with Saturn lies at about 15.14 au. Between 1.4
and 1.8 Myr, and again around 2.5 Myr, the clone lies in a region
overlapping both of these resonances, at a value of semi-major axis
very similar to that occupied today by the most unstable object
studied in our integrations, 1996 AR20 (discussed in \S 2). The
difference between this clone of 1995 SN55 and the clones of 1996 AR20
lies in the eccentricity and inclination of the objects. While 1995
SN55 is near the resonances, its eccentricity is so low that at times
it orbits entirely between Saturn and Uranus. This makes the orbit
more stable than that of 1996 AR20, which lies on a highly eccentric
orbit. In addition, the moderately high inclination of this clone
through this period ($i$ never falls below $22^\circ$ in the final 2
Myrs of the integration) helps to keep the clone stable.

When the clone is not in resonance with Jupiter and Saturn, it falls
into the 7:5 resonance with Uranus (for example, between 1.1 Myr and
1.4 Myr). The apparent untidiness in the behaviour of the orbital
elements during the final 2 Myr of the integration, given the stable
nature of the orbits occupied, is a result of the overlap between the
resonances. With whichever planet the clone is resonant at a
particular time, there will be near resonant effects from the others
involved. This may explain the rapid, small variations in $e$, $q$ and
$Q$, which are far more pronounced than those seen in clones which
occupy other resonances (for example, compare it with the behaviour
shown in Figure \ref{fig:AR20a}).

\begin{figure} 
\plottwob{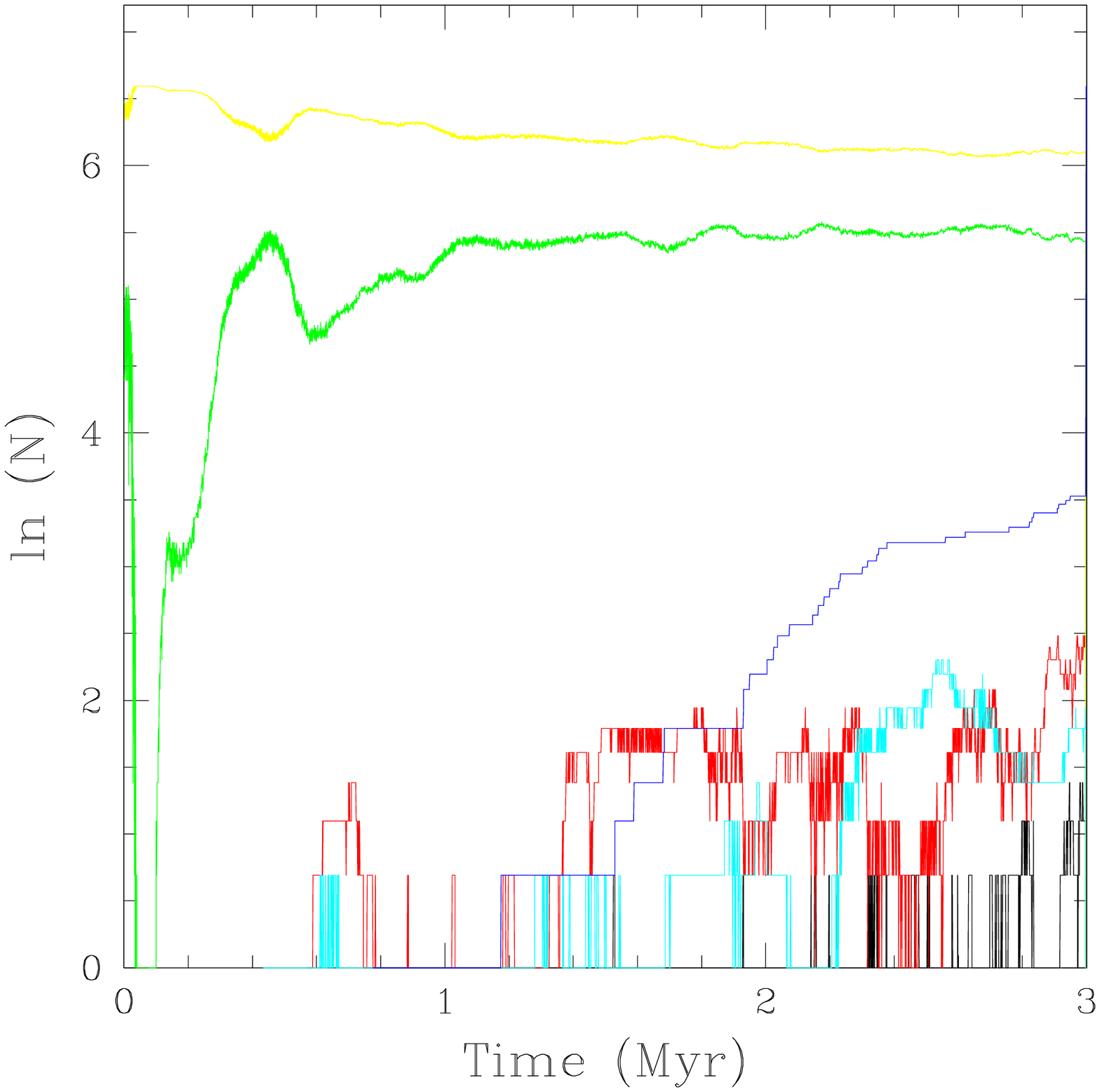}{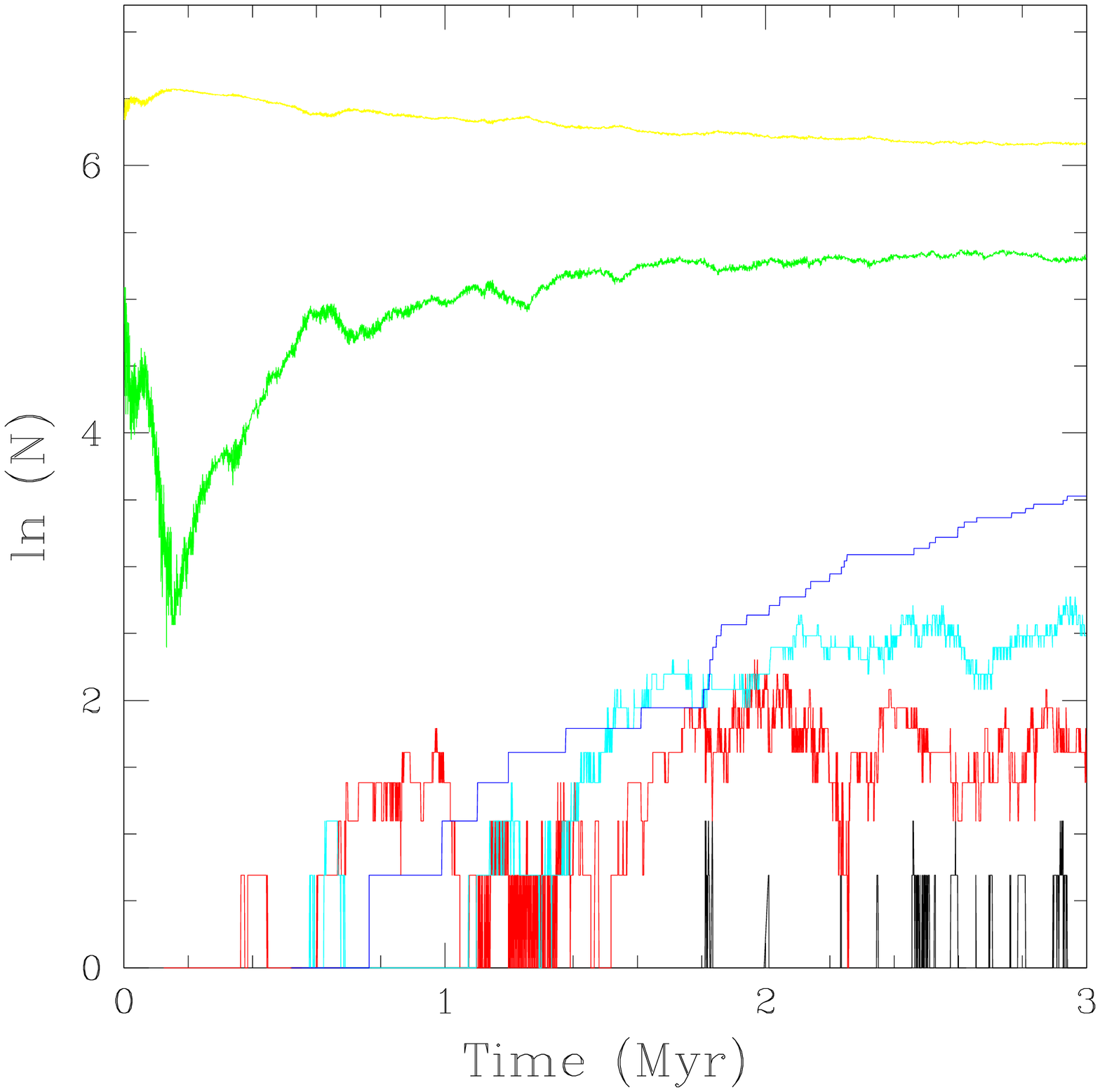} 
\caption[] {The numbers of clones of 2000 FZ35 controlled by Jupiter
(red), Saturn (green), Uranus (yellow) and Neptune (cyan), together
with the numbers of short-period comets (black), trans-Neptunian and
ejected objects (blue), plotted against time. The left (right) panel
shows the results from the forward (backward) integration.}
\label{fig:FZ53classtime}
\end{figure} 

\begin{figure}
\epsfysize=10cm \centerline{\epsfbox{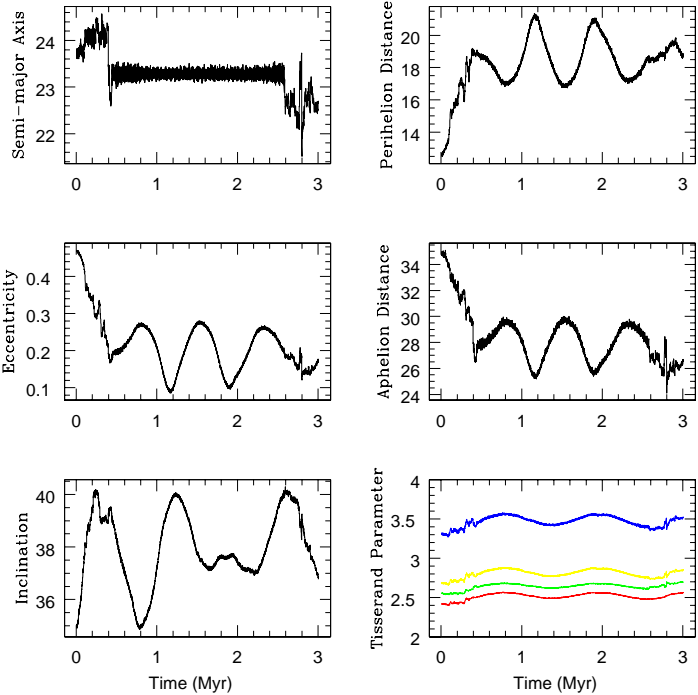}}
\caption{The behaviour of the 318th clone of 2000 FZ53 in the forward
direction. In the plot of Tisserand parameters, the value of $\TJ$ is
shown in blue, $\TS$ in yellow, $\TU$ in red and $\TN$ in green. Note
the coupled variations in eccentricity and inclination (with the
maxima of one corresponding to the minima of the other). This is
characteristic of a Kozai resonance.}
\label{fig:FZ53a}
\end{figure}
\begin{figure}
\epsfysize=10cm \centerline{\epsfbox{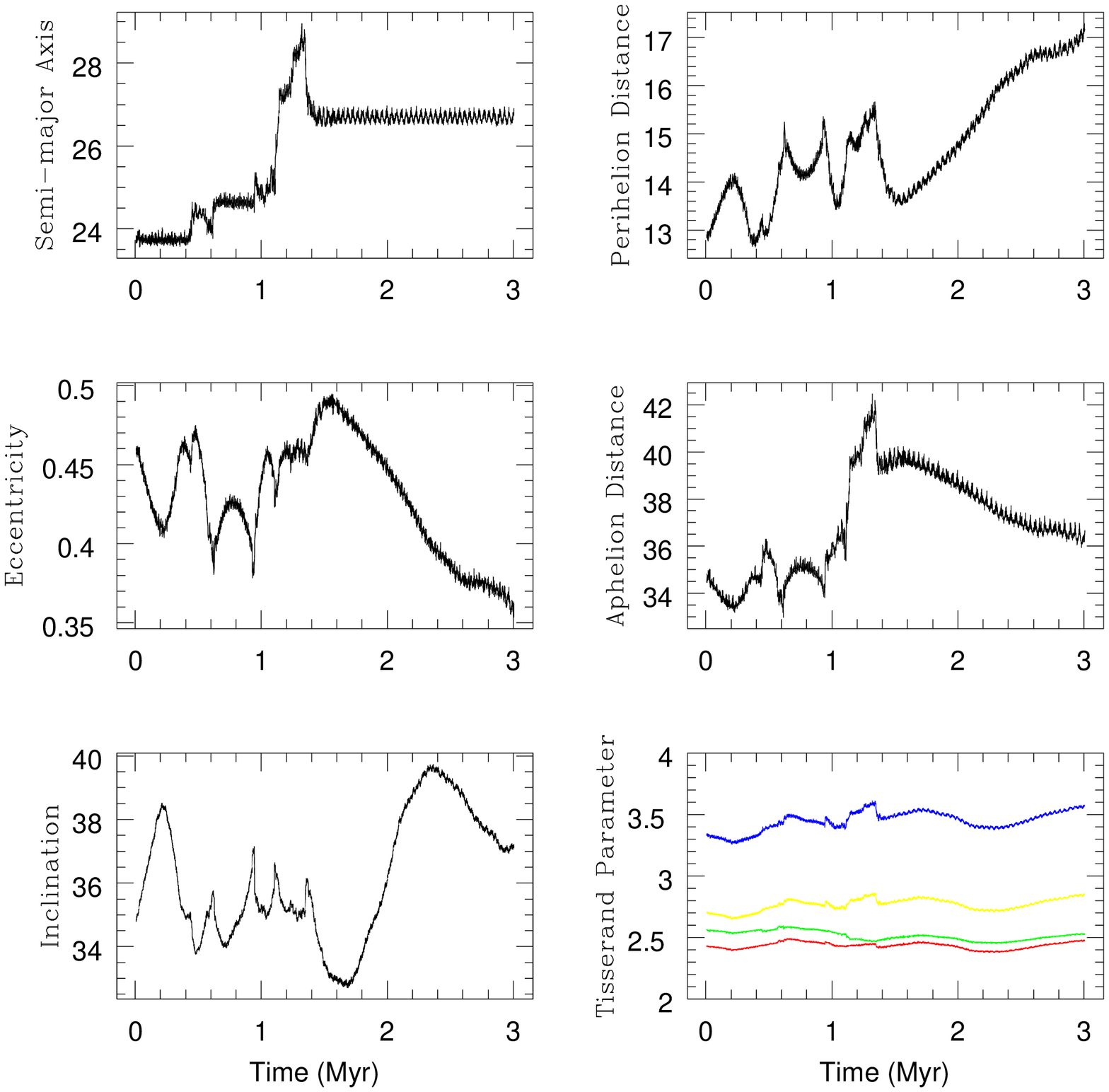}}
\caption{The behaviour of the 334th clone of 2000 FZ53 in the forward
direction. The endpoint of the clone's evolution is the 3:5 mean
motion resonance with Uranus.}
\label{fig:FZ53b}
\end{figure}

\section{Evolution of a UE object: 2000 FZ53}

At the start of the simulation, $579$ of the clones fall under Uranus'
control at perihelion and are UE objects. The remaining 150 of the 729
clones fall under Saturn's control at perihelion and are SE
objects. The distribution of the clones in $a$-$e$-$i$ space actually
straddles the boundary between UE and SE. 2000 FZ53 is the Centaur
with the longest known half-life -- approximately 26.8 Myr in the
forward and 32.3 Myr in the backward directions. It is also the object
which lies on the most highly inclined orbit of those studied ($i =
34.9^\circ$). This is a contributing factor to the longevity of the
object. Of the 729 clones in the forward direction, only 18 become
short-period comets, and a mere 5 become Earth-crossing. In the
backward direction, only 12 of the clones became short-period, with
again 5 becoming Earth-crossing.

Figure \ref{fig:FZ53classtime} shows how the overall population of
clones of 2000 FZ53 changes over time. The extreme stability of this
object is evidenced by the remarkably small number of the clones which
are ejected by the end of the simulation. After 3 Myr, less than 50 of
the initial 729 clones have been ejected in either direction of
integration. A particularly interesting feature is the extent to which
the populations of the object under the control of Uranus and of
Saturn are coupled. This is caused by the flexing of the orbit, and
hence the associated population of clones, under secular evolution.
It is a consequence of the starting configuration in which clones lie
across the boundary between SE and UE objects.  Very few of the clones
evolve inwards sufficiently to be controlled by Jupiter, or outwards
to reach Neptune.

\subsection{A Kozai Resonance}

Figure \ref{fig:FZ53a} shows the evolution of the 318th clone of 2000
FZ53. This has starting orbital elements of $a = 23.670$ au, $e =
0.469$ and $i = 34.94^\circ$. The clone spends the first 400 kyr in
various fairly stable orbits, changing occasionally through distant
encounters with Uranus and Neptune. These encounters lead to a gradual
circularization of the orbit, pulling the eccentricity down from a
value close to 0.5 at the start of the integration to a value just
below 0.2. This decrease in eccentricity causes the perihelion
distance to move outwards towards Uranus, and the aphelion distance to
fall to that of Neptune. Eventually, the clone drops into a stable 3:4
mean-motion resonance with Uranus, which it occupies for around 2.2
Myr. During this time, the clone experiences no secular changes in its
semimajor axis but there are coupled variations in eccentricity and
inclination such that $e$ is a maximum when $i$ is a minimum. This is
characteristic of an object undergoing a Kozai resonance (Kozai 1962;
Murray \& Dermott 1999), for which the Kozai integral $I_{\rm K}$
\begin{equation}
I_{\rm K} = \sqrt{1-e^2} \cos i
\end{equation}
remains constant. This can be confirmed by examining the value of
$\omega_{\rm dif} = \omega - \omega_{\rm N}$ (the difference between
the argument of pericentre for the object and Neptune). This is 
librating rather than circulating, consistent with trapping in the
Kozai resonance.

\subsection{A Mean-Motion Resonance with Uranus}

Figure \ref{fig:FZ53b} shows the evolution of the 334th clone of 2000
FZ53. This clone has initial orbital elements of $a = 23.675$ au, $e =
0.459$ and $i = 34.87^\circ$. The first 1 Myr of the evolution is
characterized by a number of protracted stays in stable orbits with
semi-major axes between 23.5 and 25 au. There is a Kozai resonance in
the first 400 kyr, during which the $e$ and $i$ of the clone vary in
the familiar coupled fashion. This is followed by a couple of small
transitions, until the clone arrives at a semi-major axis of just over
24.5 au. After the first 1 Myr, the clone experiences a series of
fairly distant encounters with both Uranus and Neptune, which change
the semi-major axis, until after 1.4 Myrs, the clone enters the 3:5
mean-motion resonance with Uranus, in which it stays until the end of
the simulation. Whilst in this resonance, the eccentricity of the
object is slowly driven down, raising the perihelion ever closer to
the orbit of Uranus.

\begin{figure} 
\plottwob{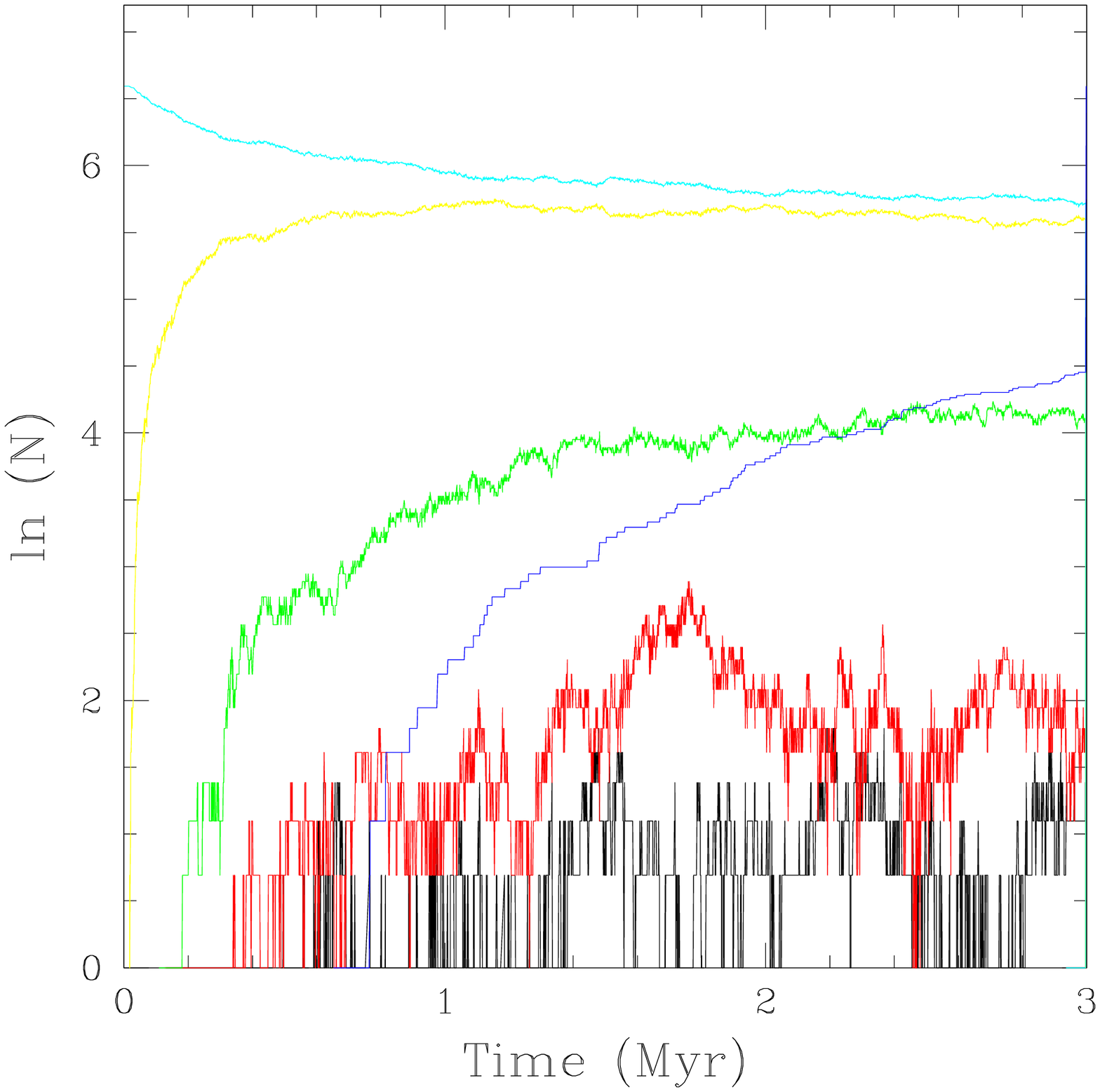}{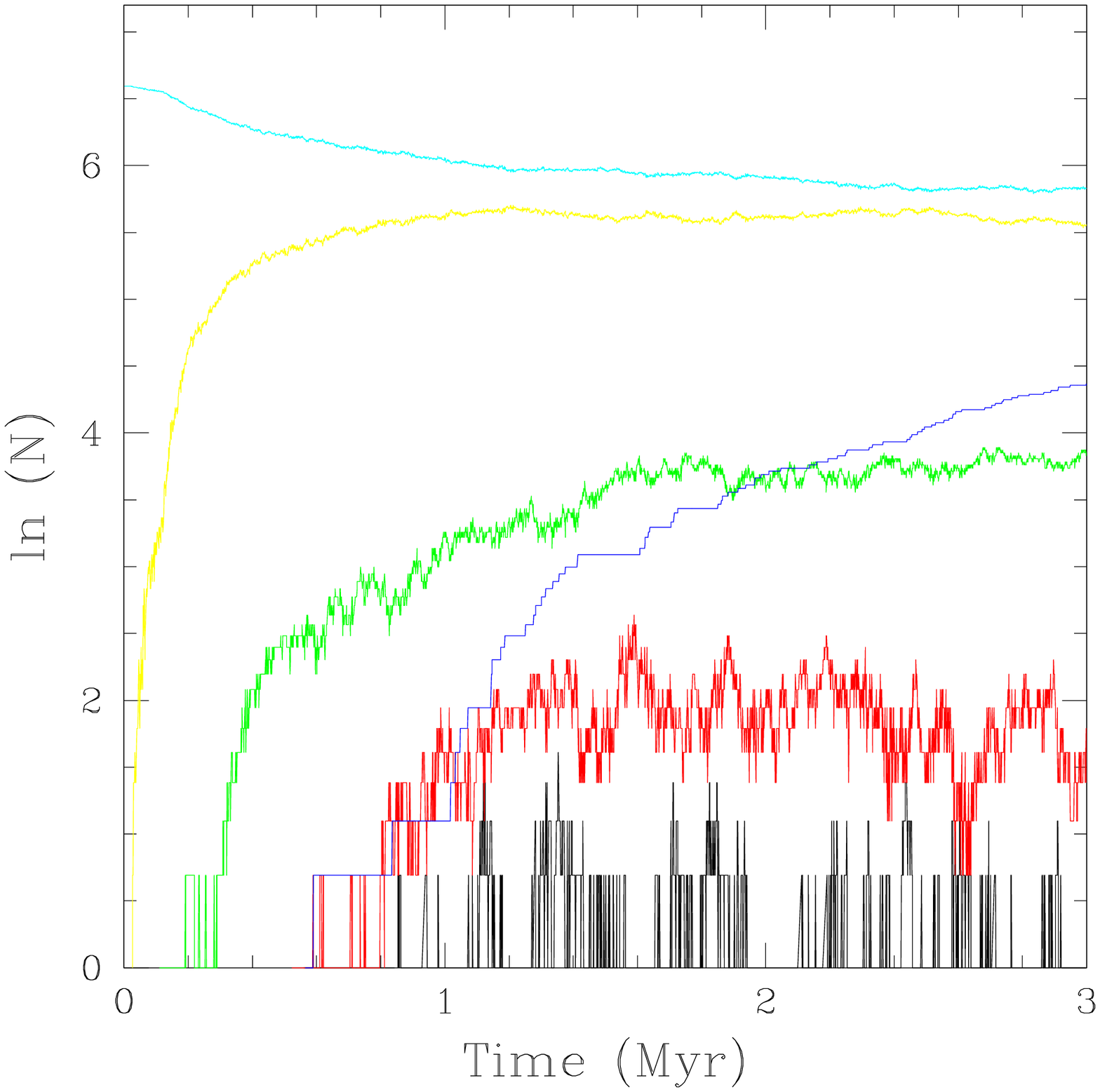} 
\caption[] {The numbers of clones of 2002 FY36 controlled by Jupiter
(red), Saturn (green), Uranus (yellow) and Neptune (cyan), together
with the numbers of short-period comets (black), trans-Neptunian and
ejected objects (blue), plotted against time. The left (right) panel
shows the results from the forward (backward) integration.}
\label{fig:FY36classtime}
\end{figure} 
\begin{figure}
\epsfysize=10cm \centerline{\epsfbox{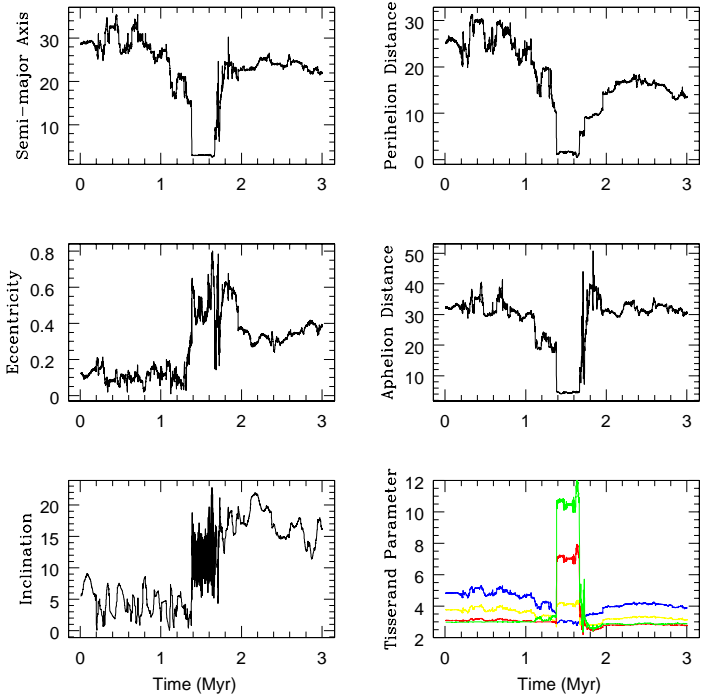}}
\caption{The evolution of the 70th clone of 2002 FY36 in the forward
direction.  direction. In the plot of Tisserand parameters, the value
of $\TU$ is shown in red and $\TN$ in green. Note that the clone
travels inwards to become an Earth-crosser (albeit briefly) before
returning to the outer Solar system.}
\label{fig:FY36a}
\end{figure}
\begin{figure}
\epsfysize=10cm \centerline{\epsfbox{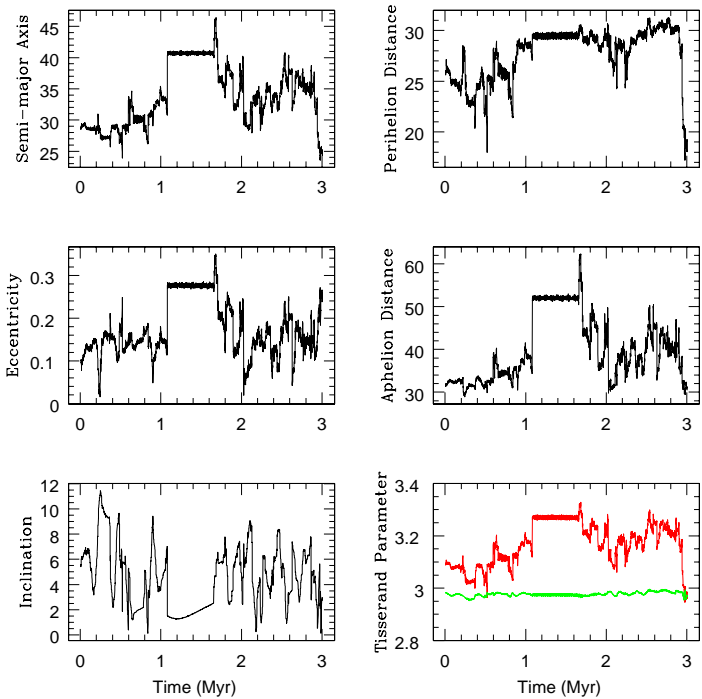}}
\caption{The evolution of the 12th clone of 2002 FY36 in the forward
direction. Note the prolonged mean motion resonance between roughly
1,1 and 1.6 Myr.}
\label{fig:FY36b}
\end{figure}

\section{Evolution of a N object: 2002 FY36}

Of all the objects studied in Paper I, the only one to be controlled
by Neptune at perihelion is 2002 FY36. This Centaur lies on a low
eccentricity orbit (in fact, it is the most circular of the seed
orbits for the integrations, with an eccentricity of 0.114). 2002 FY36
is amongst the most stable of the Centaurs, with half-lives of 12.5
Myrs and 13.5 Myrs in the forward and backward directions
respectively. In the forward simulation, only 78 of the clones of this
object become short-period objects, with 16 becoming Earth-crossing
and 35 becoming Mars-crossing. In the backward integrations these
numbers are 68, 16 and 33 respectively.  Figure
\ref{fig:FY36classtime} shows the changing populations of clones
within the simulation of 2002 FY36. The stability of the object is
evidenced both by the very slow decay of clones under Neptune's
control (around 50\% of the clones are still controlled by Neptune at
the end of the simulation), together with the very slow ejection rate
(less than 100 clones are ejected in both the forward and backward
integrations).

\subsection{A Traversal of the Solar System}

Figure \ref{fig:FY36a} shows how the 70th clone of 2002 FY36 evolved
in the forward direction. It has initial orbital elements of $a =
28.949$ au, $e = 0.124$, $i = 5.43^\circ$. This is a particularly
interesting clone since it starts the simulation purely under the
control of Neptune, and slowly works its way in through the Solar
system, becoming a short-period comet, and then works its way back out
to a reasonably stable region. Initially, the perihelion of the object
is gradually handed down to Uranus. Then, Uranus' influence (just
after the 1 Myr mark) acts to switch the perihelion and aphelion of
the object around, so that it has aphelion near Uranus and perihelion
near Saturn.  Another perihelion-aphelion interchange happens at
Saturn, handing the object down to the control of Jupiter. Jupiter
then acts almost immediately again to interchange the perihelion and
aphelion of the object, injecting it to the inner Solar system. Once
there, it resides on a series of fairly stable orbits for just over
200 kyr, before becoming Earth-crossing and then being handed back
outwards through another perihelion-aphelion interchange at
Jupiter. At around 1.75 Myr, Saturn moves the perihelion away from
Jupiter's control and moves the aphelion to Neptune's control. The
object then spends the remaining 1 Myr of the integration in an orbit
whose perihelion gets detached from Saturn by the effects of Neptune
at aphelion, and which is reasonably stable.

\subsection{A Mean-Motion Resonance with Neptune}

Figure \ref{fig:FY36b} shows the orbital evolution of the 12th clone
of 2002 FY36 in the forward direction. The initial orbital elements of
this clone were $a = 28.949$ au, $e = 0.104$ and $i =
5.37^\circ$. This clone is captured into a resonance when its
semi-major axis is just over 40 au. This is close to both the 1:3
resonance with Uranus and the 7:11 resonance with Neptune. Whilst in
the resonance, the eccentricity (and hence the perihelion $q$ and
aphelion $Q$ distance) of the clone is remarkably stable, as, to a
lesser extent, is the inclination. This is likely an artefact of the
lack of perturbing objects beyond Neptune. In practice, the effect of
perturbations of massive bodies in the Edgeworth-Kuiper belt is likely
to decouple such objects from Neptune altogether (and obviously, given
that such behaviour is time-reversible, lead to the injection of fresh
objects from such areas).

\section{Conclusions}

We have presented 3 Myr integrations of the orbits of clones of five
Centaurs -- namely, 1996 AR20, Chiron, 1995 SN55, 2000 FZ53, and 2002
FY36. At the start of the integrations, there is one object with
perihelion under the control of Jupiter (1996 AR20), two under the
control of Saturn (Chiron and 1995 SN55), and one each under the
control of Uranus (2000 FZ53) and Neptune (2002 FY36). As the
simulation evolves, clones of the Centaurs diffuse throughout the
Solar system. This is illustrated by the behaviour of the number of
clones controlled by each planet over time. The examples presented
here are just a small number from the grand total of 23\,328 Centaur
orbit integrations carried out for our statistical analysis (Horner et
al. 2004, henceforth Paper I).

There are a number of generic patterns of behaviour identified from
the simulations and illustrated by our examples. Every Centaur
produces some clones which show short-period cometary activity during
the 3 Myr evolution. Chiron has over $60 \%$ of its clones becoming
short-period objects, whilst 1995 SN55 has over $35 \%$. Clones of
these Centaurs typically make numerous close approaches to Jupiter. At
the other extreme, 2000 FZ53 has $\sim 2 \%$ of its clones becoming
short-period objects. It has been argued that the injection of a large
Centaur like Chiron or 1995 SN55 into the inner Solar system will
produce major biological and climatic trauma on the Earth (e.g., Hahn
\& Bailey 1990, Bailey, Clube \& Napier 1990). If a clone becomes a
short-period object, then it is likely to have repeated bursts of
short-period activity -- on average $\sim 30$ or so in our
simulations.  Chiron is likely to be such a serial offender, as its
blue colours probably point to a spell of short-period cometary
activity in the recent past.  Further such forays into the inner Solar
system may well take place in its future.

About $20 \%$ of the clones which become short-period comets then go
on to become Earth-crossing. The idea that cometary bodies may
populate the Earth-crossing asteroid families can be traced back to
\"Opik (1963). This is not the only source of Near-Earth objects, as
asteroids in the Main Belt lying near the 3:1 resonance with Jupiter
can also be transferred to Earth-crossing orbits (e.g., Wisdom 1983,
1985). Estimates of the fraction of near-Earth objects (NEOs)
emanating from the Main Belt vary between $40 \%$ (Wetherill 1988) and
$\gta 80\%$ (Ipatov 1989, Bottke et al. 2002). Nonetheless, the
evidence that some dead comets become NEOs is strong. For example, the
near-Earth asteroids 2201 Oljato and 3200 Phaethon are convincing
cometary candidates, either on the grounds of surface composition
(McFadden et al. 1984) or of links to known meteor showers (Whipple
1983). Our calculations suggest that one Centaur becomes
Earth-crossing for the first time approximately every $\sim 880$ yrs
(see Paper I). The example presented in this paper is a possible
evolutionary pathway for the largest known Centaur 1995 SN55, which
has a diameter between 170 and 380 km. This emphasizes the possible
dangers of objects emanating from the Centaur region -- Centaurs are
typically larger and more massive than asteroids. Even if they are not
the major contributor to the Near-Earth population in numbers, their
contribution to the high-mass end is likely to be overwhelming.

A number of our Centaur clones become trapped at 1:1 mean motion
resonances around the giant planets. Here, we presented an example of
a clone of 1996 AR20 which spends 0.5 Myr in a tadpole orbit around
the 1:1 resonance with Jupiter.  Studies of the origin of the Jovian
Trojans usually assume that they are primordial. During the early
stages of the formation of Jupiter, planetesimals are trapped into the
changing gravitational field around the growing planet. Mutual
collisions or energy losses due to gas drag may drive trapped
planetesimals deeper into stable Trojan orbits (e.g., Shoemaker et
al. 1989, Marzari et al. 2003). Based on our orbital integrations, an
entirely new supply route is possibly, namely the capture of
Centaurs. This may be tested by looking for out-gassing from Jovian
Trojans, as any recently captured Centaurs may still contain
volatiles.  The supply route works for the other giant planets as
well.  An example of a clone of Nessus captured into a horseshoe orbit
around the 1:1 resonance with Uranus will be presented elsewhere.
This suggests that the Trojan populations of all the giant planets
may be partly sustained by the flux of Centaurs. 

The net flux of the Centaur population is inward, as the primary
source is the Edgeworth-Kuiper belt whilst Jupiter tends to eject the
objects from the Solar system over the course of time. Nonetheless,
examples of outward migration of individual clones often occur in the
simulations, as illustrated by particular clones of Chiron, 1995 SN55
and 2002 FY36 in this paper. The former is particularly remarkable as
it moves all the way in to Earth-crossing, before moving all the way
back out to beyond Saturn. A burst of short-period cometary activity
is followed by a return to the domain of the Centaurs. Such repeated
traversals of the Solar system are a defining characteristic of the
Centaur population, which is therefore expected to include objects
encompassing a wide range of differing physical and dynamical
characteristics.

\section*{Acknowledgements}
JH thanks PPARC for the award of a studentship. Research at Armagh
Observatory is supported by the Northern Ireland Department of
Culture, Arts and Leisure.


\label{lastpage}
\end{document}